\documentclass[12pt]{iopart}
\usepackage{epsf}
\begin{document}
\newcommand{\EQ}{\begin{equation}}
\newcommand{\EN}{\end{equation}}
\newcommand{\EQA}{\begin{eqnarray}}
\newcommand{\EQN}{\end{eqnarray}}
\newcommand{\EQAN}{\begin{eqnarray*}}
\newcommand{\EQNN}{\end{eqnarray*}}

\newcommand{\Sp}{{\rm Sp}}
\renewcommand{\theequation}{\arabic{section}.\arabic{equation}}
\begin{flushright}
hep-th/9908153\\
UT-KOMABA/99-12\\
August 1999
\end{flushright}
%


\title[Generalized Conformal Symmetry  and 
Oblique AdS/CFT Correspondence . . .
]{Generalized Conformal Symmetry and 
Oblique AdS/CFT Correspondence for Matrix Theory
}

\author{ Tamiaki Yoneya\footnote{e-mail address: tam@hep1.c.u-tokyo.ac.jp}}

\address{
Institute of Physics, 
University of Tokyo, Komaba, Tokyo 153-8912, Japan}

\begin{abstract}
The large $N$ behavior of  Matrix theory is discussed on the basis of the previously proposed generalized conformal symmetry.  The concept of `oblique' AdS/CFT 
correspondence, in which the conformal 
symmetry involves both the space-time 
coordinates and the string coupling constant, is proposed. 
Based on  the explicit  predictions for
 two-point correlators,  
  possible implications  for 
the Matrix-theory conjecture are discussed.  
 
\end{abstract}




\vspace{0.3cm}
\noindent
The Matrix-theory conjecture \cite{bfss} 
requires us to investigate the dynamics 
of D-particles described by the supersymmetric Yang-Mills 
matrix quantum mechanics in the large $N$ limit. 
Unfortunately,  very little is known 
as for the relevant large $N$ behaviors of the 
matrix quantum mechanics. 
In the present talk, I shall discuss  the large $N$ limit 
of Matrix theory 
by extending the AdS/CFT correspondence 
to the matrix quantum mechanics using the 
previously proposed generalized conformal symmetry as a guide.    
  
The paper is organized into three parts. In the first part, 
I begin by briefly recalling the Matrix-theory conjecture 
and review the so-called DLCQ interpretation at finite $N$. 
The latter interpretation will be used as an 
intermediate step for our later arguments. 
Then in the second part, after a brief 
discussion on  the 
generalized conformal symmetry \cite{jy, jky}  from the 
point of view of the AdS/CFT correspondence, I 
introduce the notion of `oblique' AdS/CFT correspondence 
for nonconformal D0-branes.  
 In the third part, 
I discuss the results of the harmonic analysis 
of supergravity fluctuations around the D-particle 
background and its predictions for the 
two-point corrrelators of Matrix theory 
operators in the large $N$ limit, based on our  
recent work  \cite{sekiyo} 
which contains, to my knowledge, 
the first extensive computation of the 
correlators for {\it dilatonic} case based on the 
AdS/CFT correspondence.  I then propose to interpret the anomalous large $N$  scaling behavior 
  found from this  analysis as an indication of a screening mechanism  
which may reconcile the holographic growth 
of the transverse size with 11 dimensional boost invariance.  

\section{Matrix-theory conjecture and the DLCQ 
interpretation}

The basic assumption of Matrix theory \cite{bfss} is that the 
low-energy effective super Yang-Mills theory of D-particles 
in type IIA superstring theory is the 
exact description of `M-theory' in the infinite-momentum limit where the total 11th dimensional momentum 
$P_{10}$ becomes infinite: 
\EQ
P_{10}=N/R \rightarrow \infty ,
\EN
where $R=g_s\ell_s$ is 
the radius of the 11th dimension, which is 
compactified to a circle, and $N$ is the number of 
D-particles.  The D-particles are now interpreted as 
partons as the basic constituents of M-theory.  The effective 
action is 
\EQ
S =\int dt \, \Tr 
\Bigl( {1\over 2g_s\ell_s} D_t X_i D_t X_i + i \theta D_t \theta 
+{1 \over 4g_s\ell_s^5} [X_i, X_j]^2 -
 .... ) .
\label{action}
\EN
In the infinite-momentum limit, the 11th momentum and 
the Hamiltonian can be identified with the light-like 
momentum and the light-like Hamiltonian, respectively. 
\EQ
P_{10}\rightarrow P_{-}=N/R ,
, \quad 
H\rightarrow P^{+}=-2P^- =R\Tr h={N\over P_-}\Tr \, h
\label{Hamilton} ,
\EN
\EQ
h={1\over 2}\Pi^2  - {1\over 4\ell_P^6}[X^i, X^j]^2 
+ {1\over 2\ell_P^3}
[\theta_{\alpha},[X^k, \theta_{\beta}]]\gamma^k _{\alpha\beta} ,
\label{hamiltonian}
\EN
where we have introduced the 11 dimensional 
Planck length $\ell_P=g_s^{1/3}\ell_s$. 
For any finite  and fixed $R$, 
the infinite-momentum limit requires 
the large $N$ limit, $N\rightarrow \infty$.  
The (super) Galilean symmetry of this system is indeed 
consistent with the infinite-momentum frame interpretation. 

Obviously,  the infinite-momentum limit 
can also be achieved for fixed $N$ if we take the limit 
of small compactification circle, $R\rightarrow 0$, 
namely the type IIA limit. 
If we boost back to a {\it finite} $P_-'=N/R'$, this defines 
a theory in which the system is effectively 
compactified along the light-like direction $x^-
\sim x^-+ 2\pi R'$.  This is nothing but the 
discrete light-cone interpretation (DLCQ), 
proposed by Susskind \cite{suss} and elucidated in \cite{seibergsen}.  
The basic assumption for this interpretation 
is  that the dynamics is invariant under the 
boost along the 11th direction for finite and fixed $N$ 
in contrast to the original BFSS conjecture 
which assumes that the boost is associated with 
a change of $N$ for fixed $R$.   
In fact, if we fix the 11 dimensional Planck length, 
the Hamiltonian and $P_-$ transform as desired, 
$P^- \rightarrow \rho P^-$  
under the change, 
$R\rightarrow \rho R \leftrightarrow P_-\rightarrow \rho^{-1}P_-$,  of the compactification radius.   
This may be called as the `kinematical' boost 
transformation, contrasting to the `dynamical' 
boost of the BFSS conjecture.  
Usually, the limit $R\rightarrow 0 \, \, (R\ll \ell_P) $ which corresponds to the weak-coupling limit $g_s\rightarrow 0$ is 
regarded as justifying the matrix model, since the 
characteristic length scale $\ell\sim \ell_P$ of the model 
is now much smaller $\ell_P \ll \ell_s$ than the string scale, and hence 
the massive string modes of the open strings 
stretched among D-particles are decoupled 
as long as we are interested only in the energy range 
much smaller than the characteristic string unit 
$E (\sim P^+) \ll \ell_s^{-1}$.   The two conditions, $\ell\sim \ell_P$ 
and $E \ll \ell_s^{-1}$, which might look naively 
contradictory to each other, can be compatible since 
now the effective mass $m \sim 1/g_s\ell_s$ in  
10 dimensional space-time is very heavy and the characteristic 
velocity of D-particles is given by $v\sim g_s^{2/3}$. 

However, it should be emphasized that the DLCQ interpretation cannot be regarded as a `proof' that the 
matrix model is consistent with supergravity in the 
long-distance limit.  In the limit of small compactification radius, 11 dimensional supergravity reduces to 10 dimensional type IIA supergravity. Only natural 
justification of type IIA supergravity relying upon the dimensional arguments is the zero-slope limit in which 
$\ell \gg \ell_s$ by keeping the 11 dimensional Planck length 
$\ell_{10}=g_s^{1/4}\ell_s$ fixed.  This limit clearly 
exceeds the range of the DLCQ region $\ell \ll \ell_s$.  
Note that the low-energy limit $\ell \gg \ell_P$ in the sense 
of 11 dimensions can be used as a criterion {\it only} in the 
{\it de}compactification ({\it i.e},  strong coupling) 
limit $R\rightarrow \infty$ for fixed $\ell_P$. 

In spite of this apparent disagreement on the 
ranges of validity,  
the leading low-velocity expansions for the 
scattering phase shifts for the matrix model 
and supergravity are known to give precisely 
the same results at long-distance regime $\ell \gg \ell_P$ in 
lower order  perturbation theory with respect to $g_s$.  
The supersymmetric nonrenormalization theorem \cite{nonren} 
is responsible for this result at least to the first nontrivial 
order.  It is not clear whether the supersymmetry is sufficient to explain a 
much more nontrivial result \cite{okawayoneya} at two-loop 
order which contains the nonlinear self-interaction 
of graviton. An important task in Matrix theory is to clarify to what extent the coincidence between the matrix model for finite $N$ and 
the supergravity in the DLCQ limit is valid, 
and, if the coincidence stops at some point, where and how it occurs. 
This is not, however,  the issue on which I would like to focus 
in the present talk. For a previous review, see {\it e.g.} \cite{yonishi}.   Let us return to the large $N$ limit.  

In contrast to the DLCQ interpretation,  the 
Matrix-theory conjecture requires that boost transformation 
in going to the infinite-momentum frame is equivalent to 
taking the large $N$ limit with fixed $R$. Since $P^{+}P^{-}=
-N\Tr h /2$ has to be boost-invariant, the whole 
nontrivial spectrum must then be contained in the 
region where the spectrum of the operator $\Tr h$ 
scales as $O(1/N)$. Namely, the Hamiltonian $P^{-}$ must be scaled as $O(R/N)$. If we study the correlators of the 
theory instead of the spectrum directly, this amounts to 
taking the scaling limit with respect to time as 
$t \rightarrow Nt$.  In view of the analogy 
with light-cone formulation of membranes, it is not unreasonable to suppose that an appropriate large $N$ 
limit of the matrix model captures the full structure of type IIA/M theory.  As we shall discuss shortly, 
the large $N$ limit indeed enables us  to 
go beyond the DLCQ region, namely, go to the 
region of length scales much larger than 
the string scale. 

One of the puzzles related to this conjecture is that 
the holographic property which should be satisfied as 
a proper quantum theory of gravity requires that the 
size of the system with respect to the transverse directions 
must grow indefinitely in the large $N$ boost transformation. 
The reason is that the boost transformation increases the 
number of partons.  Since the information carried by 
a single parton is expected to be coded in a transverse volume 
at most of Planck size, the transverse volume of the total system must 
grow at least as fast as $N$ which means that the 
transverse size is at least of order $O(N^{1/9})$ 
in the limit $N\rightarrow \infty$.   Thus a crucial issue in 
studying the large $N$ behavior of Matrix theory 
is how to reconcile these   
properties with the boost invariance.  
In the  light-cone formulation of 
string theory, we parametrize the 
strings such that the density of the 
light-like momentum is uniform along the string and also is  constant with respect to 
the light-cone time.    The number of string bits 
is then proportional to the light-like momentum.  
However no violation of Lorentz invariance occurs 
in the final results of scattering amplitudes.  
In Matrix theory, similarly, 
  there must be 
some mechanism by which the apparent growth of 
the transverse size could be compatible with 
11 dimensional Lorentz invariance .

\section{Generalized conformal symmetry and 
oblique AdS/CFT correspondence}

Let us now consider the following question: 
Can we extract any nontrivial information 
on the large $N$ limit of Matrix theory 
from the AdS/CFT type correspondence? 
This problem must have been touched upon by 
many authors  from different 
perspectives.  Due to the limited space, I can only mention a few  \cite{polchinski} \cite{townsend} 
which are closely related to the present talk.   
For other related works, I recommend the readers to consult 
references in these cited works.  However, to my knowledge, 
no concrete results have been reported on the 
behavior of the correlation functions of Matrix 
theory along this direction.  In our
 previous works \cite{jy, jky}, 
we have proposed to approach the problem 
from the viewpoint of a generalized conformal symmetry.  

In the usual AdS/CFT correspondence, the 
existence of conformal symmetry plays 
crucial roles. The classification of the spectrum 
with respect to the conformal symmetry on both sides of 
bulk and boundary theory and agreement between 
them are the strongest piece of evidence for the 
correspondence.  Furthermore, the 
conjectured correspondence of the correlators 
between the bulk theory 
and the  conformal field theory at the boundary of the near-horizon region again relies upon the 
conformal symmetry of the bulk theory 
in the whole near-horizon region. 
Therefore it is natural to 
seek the possible generalization of the conformal symmetry 
for non-conformal branes in extending the correspondence 
to non-conformal D-branes. 

Let us start by examining the structure of the classical 
D0-solution: 
\EQ
ds_{10}^2 = -e^{-2\tilde{\phi}/3}dt^2 + e^{2\tilde{\phi}/3}dx_i^2 , \quad 
e^{\phi}= g_s e^{\tilde{\phi}}
\label{dilaton}
\EN
\EQ
e^{\tilde{\phi}} =\bigl( 1 +{q\over r^7}\bigr)^{3/4} , 
\quad 
A_0=-{1\over g_s}\bigl( 
{1\over 1+{q\over r^7}}-1\bigr)
\bigr) , 
\EN
where the charge $q$ is given by 
$
q=60\pi^3 (\alpha')^{7/2} g_sN .
$
In the near horizon limit $q/r^7\gg 1$, 
the factor $1+q/r^7$ is replaced by $q/r^7$ and the metric 
is rewritten as 
\EQ
ds_{10}^2=-{r^2\over \rho^2} dt^2 +{\rho^2\over r^2}(dr^2 
+r^2 d\Omega_8^2) , \quad 
\rho=\rho(r) = \Bigl({q\over r^3}\Bigr)^{1/4} . 
\EN
We can check that 
the metric, dilaton and the 1-form $A_0dt$ are all invariant 
under the scale and the special conformal transformations 
if they are accompanied by the transformations  
of the string coupling as 
\EQ
r\rightarrow \lambda r, \, 
t \rightarrow \lambda^{-1}t ,\, 
g_s \rightarrow \lambda^3 g_s ,
\label{sugrascale}
\EN
 \EQ
\delta_K t = - \epsilon (t^2 +{2q\over 5 r^5}) , 
 \quad \delta_K r =2 \epsilon tr ,
\quad \delta_K g_s=6 \epsilon t g_s ,
\label{sugraspecial}
\EN
which together with time translation form an 
$SO(1,2)$ algebra.  An important feature of 
this generalized conformal symmetry is that 
the would-be AdS radius $\rho$ as a function 
of $r$ is invariant under the transformation. 
 Namely, although the 
background space-time is {\it not}  AdS$\times S^8$, 
it behaves almost like that,  since the $r$-dependent 
radius $\rho(r) $ is invariant. 

Furthermore, the same generalized symmetry is satisfied 
for Matrix-theory lagrangian:
\EQ
X_i \rightarrow \lambda X_i, \, \, t\rightarrow \lambda^{-1}t, \, \, 
g_s\rightarrow \lambda^3 g_s , 
\EN
\EQ
\delta_K X_i = 2 \epsilon  t X_i ,   
 \delta_K t =-  \epsilon t^2 , \, \, \delta_K g_s =6\epsilon  t g_s . 
\label{eq29}
\EN
The difference between $\delta_Kt$   
in (\ref{sugraspecial}) and (\ref{eq29}) 
has the same origin as in the usual case of D3-brane: 
The mechanism how the additional term 
${2q\over 5 r^5}$ in (\ref{eq29}) emerges in the bulk theory 
was clarified in refs. \cite{jky} for general case 
of D$p$ -branes from the point of view of matrix models, 
namely, from the boundary theory. I also remark that 
 this generalization 
of conformal symmetry has been motivated at a deeper level  
by the space-time uncertainty principle \cite{yo} as has been discussed in \cite{jy} in detail.  

Another comment which is perhaps worthwhile to make here is that the generalized conformal symmetry 
is regarded as the underlying symmetry for the 
DLCQ interpretation.  We are free to change the 
engineering scales. Thus, if one wants to 
keep the numerical value of the transverse 
dimensions, we perform a rescaling $t\rightarrow \lambda^{-1} t,  
X_i\rightarrow \lambda^{-1}X_i ,\ell_s\rightarrow 
\lambda^{-1}\ell_s $ simultaneously with the 
generalized scaling transformation leading to 
the scaling 
$t\rightarrow \lambda^{-2} t, 
  X_i\rightarrow X_i, R\rightarrow \lambda^2 R$ and $\ell_P \rightarrow \ell_P$
which is equivalent with the kinematical boost 
transformation. Alternatively, 
one might want to keep the numerical value of 
time or energy by making a rescaling 
$t\rightarrow \lambda t,\quad X_i\rightarrow \lambda X_i,   
\ell_s \rightarrow \lambda \ell_s$, 
leading to the scaling 
$t\rightarrow t, R\rightarrow \lambda^4 R, X_i \rightarrow \lambda^2 X_i, \ell_P 
\rightarrow \lambda^2\ell_P$ and $\ell_s \rightarrow \lambda \ell_s$, which is in fact equivalent to the 
`tilde' transformation utilized in \cite{seibergsen}.  
Note that 
although the second case makes the 
string length $\ell_s$ small by assuming small $\lambda$, 
the length scale of transverse directions smaller than the 
string scale is always sent to even smaller length scale 
$(<\lambda^2\ell_s)$.   

In view of these symmetry properties, it is quite natural 
to suppose that the similar correspondence 
between supergravity and Yang-Mills matrix 
quantum mechanics is valid as in the typical case 
of D3-brane between supergravity and 4 dimensional super Yang-MIlls theory. 
We now examine the conditions \cite{maldacena} \cite{itzak} for the validity of 
the correspondence.  For comparison, we indicate the 
corresponding conditions for the case of D3-brane in 
parentheses.  We neglect numerical 
coefficients in writing down these conditions. 
\begin{itemize}
\item Near horizon condition : 
\EQ
r \ll \rho(r) \rightarrow r \ll (g_sN)^{1/7}\ell_s ,  \quad (r \ll (g_sN)^{1/4}\ell_s : {\rm D3})
\label{nhcondition}
\EN 

\item Small curvature condition :
\EQ
\rho(r) \gg \ell_s  
\rightarrow r \ll (g_sN)^{1/3}\ell_s
,\quad ((g_sN)^{1/4}\ell_s \gg \ell_s : {\rm D3})
\label{smccondition}
\EN

\item Small string-coupling condition:
\EQ
g_s\e^{\tilde{\phi}} =\e^{\phi} \ll 1 
\rightarrow (g_sN)^{1/3}N^{-4/21}\ell_s
\ll r
. \quad (g_s \ll 1 : {\rm D3})
\label{sstrccondition}
\EN
\end{itemize}

\noindent
In both cases of 
D0 and D3, the near horizon 
conditions (\ref{nhcondition}) are {\it by definition} not invariant 
under the (generalized) conformal 
transformation, while the other two conditions 
(\ref{smccondition}) and (\ref{sstrccondition}) 
are invariant.  The former indicates that the 
boundary of the near-horizon or conformal 
region should be assumed at $r\sim q^{1/7} 
\quad (r\sim (g_sN)^{1/4} : {\rm D3}) $.  
Note that in the case of ordinary AdS/CFT correspondence, 
both the near-horizon 
condition and the small curvature conditions are 
characterized by a single scale $(g_sN)^{1/4}\ell_s=
\ell_{10}N^{1/4}$, 
while in the present case they are different. 

Both the  difference and the common features 
in the above conditions  for D0 and D3 are best illustrated by Figures 1 and 2 below.  
\begin{figure}[hbtp]
\begin{center}
\begin{picture}(300,190)
\put(-90,-72){\epsfxsize 293pt 
\epsfbox{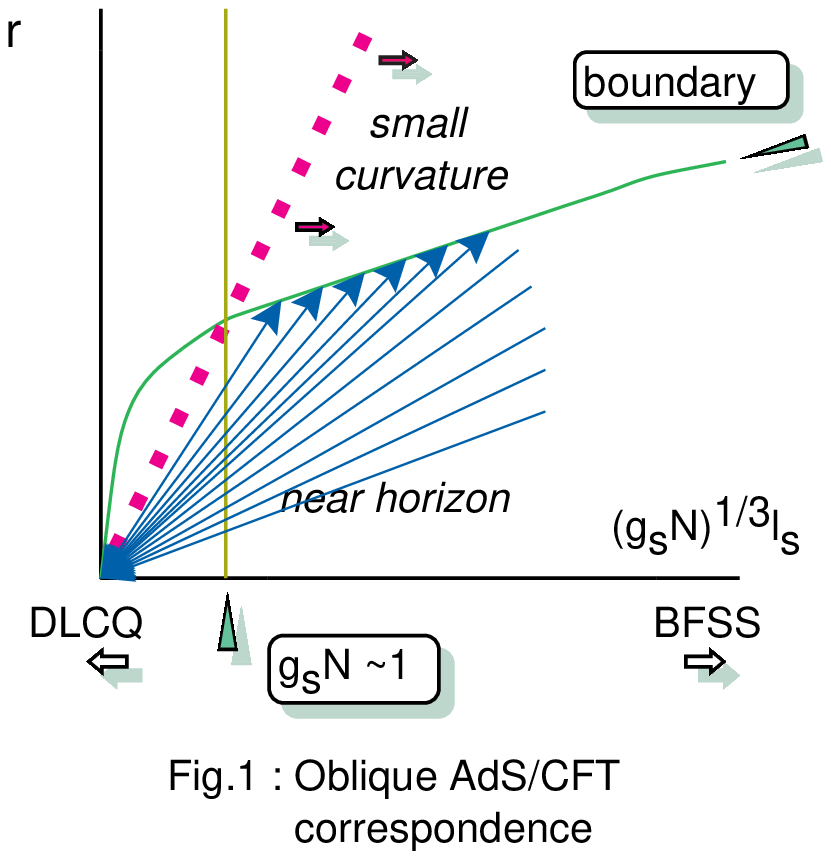}}
\end{picture}
\end{center}
\end{figure}
\begin{figure}[htbp]
\begin{center}
\begin{picture}(270,10)
\put(155,0){\epsfxsize 236pt  
\epsfbox{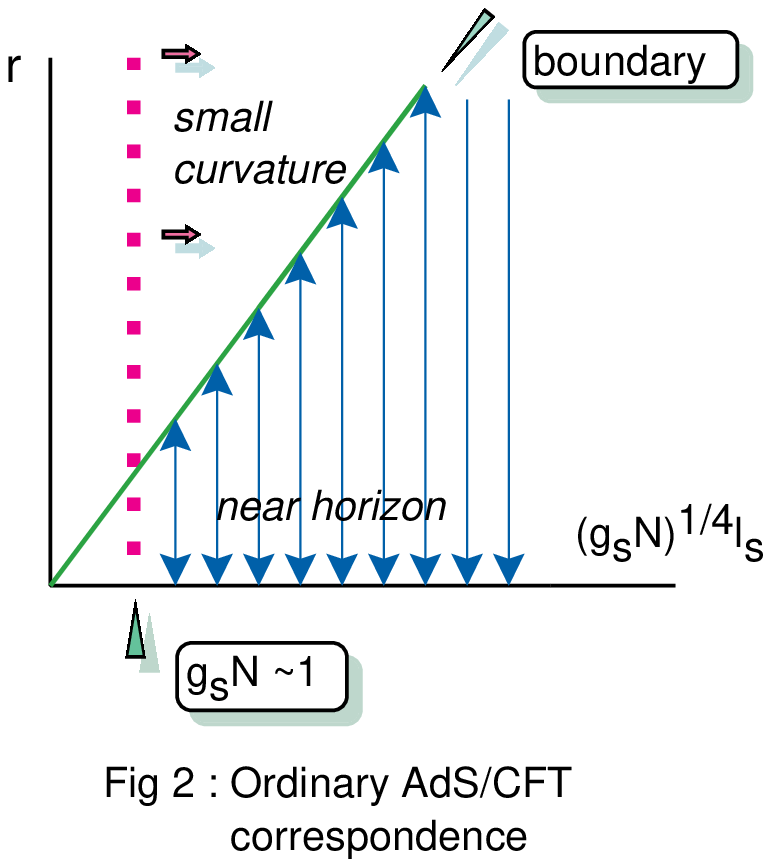}}
\end{picture}
\end{center}
\end{figure}
\vspace{-0.8cm}
In Fig. 1,  the lines 
representing the near-horizon boundary and the small-curvature boundary (dashed), 
as well as the 
flow lines (with arrows) of the generalized conformal transformation,  
are tilted comparing to the corresponding lines 
of Fig. 2, while the topologies of those lines 
are essentially the same for $r>0$.  
Note that the condition of small string coupling is 
automatically satisfied for finite $r>0 $ in the 't Hooft limit by keeping $g_sN$ fixed. 
In the oblique case, contrary to the ordinary case, 
the near horizon region with 
small effective curvature 
contains the DLCQ region $g_sN < 1$.  This 
is due to the $r$ dependence of the 
quasi AdS size $\rho(r)$. However, 
this does not mean that the correspondence 
is valid when restricted in the weak coupling region, since 
by the generalized conformal transformation 
even the DLCQ region is sent  
to the strong coupling region $g_sN > 1$ 
at the near-horizon boundary. €

On the other hand, the limit of large 
coupling 't Hooft $g_sN \rightarrow \infty$ 
enables  us to approach the BFSS region. Here, it is 
crucial to realize that the oblique AdS/CFT correspondence 
is limited at the boundary of the near horizon condition 
which is at the distance of order $ r \sim q^{1/7} \sim (g_sN)^{1/7}\ell_s$, 
just like that the ordinary AdS/CFT correspondence 
for D3 case is limited at the boundary $r\sim (g_sN)^{1/4}
\ell_s$.  It is natural to suppose that these 
distances are the infrared cutoffs for  the corresponding 
Yang-Mills theory with respect to 
the range of values of transverse 
coordinates, namely, the cutoffs for the magnitude  of the 
Higgs fields $X_i$.

\section{Predictions for 2-point functions 
and anomalous large $N$ scaling behavior}

Armed by these considerations, we are now ready to 
discuss the large $N$ behavior of the 
two-point correlators. 
The strategy for doing this is the following. 
First step: We establish the correspondence between 
the supergravity fluctuations and the matrix-model 
operators relying upon the generalized conformal 
symmetry.  Namely, we derive the two-point 
correlators assuming the now familiar conjecture 
that the supergravity action is the generating functional 
of the matrix model correlators.  We will follow the 
prescription of \cite{gpk}, which assumes the 
boundary at $r\sim q^{1/7} \, \, (r\sim (g_sN)^{1/4}\ell_s : D3)$ and extracts the universal part of the 
correlators as the nonanalytically (in momentum space) behaved part.  This is more convenient for us 
than the more formal prescription adopted in 
\cite{witten}.  We also note that the 
singularity of the metric and the dilaton 
essentially cancels at the horizon and the 
behavior of the kinetic radial term of the effective action at the horizon is not much different from the 
D3 case:
{\it e. g.}
$
\partial_{\mu}\sqrt{g}e^{-2\phi}g^{\mu\nu}
\partial_{\nu} \propto     r^8
(\partial_r^2 +8{1\over r}\partial_r -{q\over r^7}\partial_t^2) 
$ as $r\rightarrow 0$ for the 
scalar field. 
Hence the 
singularity does not cause any harm for our program.   
As remarked previously, the 
generalized scaling transformation is 
equivalent to the kinematical boost transformation.  
Thus the first step amounts to utilizing the DLCQ 
interpretation of the matrix model.  
Since the range of validity for the oblique AdS/CFT 
correspondence  contains the 
DLCQ region ($g_sN \rightarrow 0$) and BFSS region 
($g_sN \rightarrow \infty $)as two 
opposite extreme limits (see Fig. 1), we can expect that 
the behavior of the correlators  should also be 
consistent with the BFSS interpretation, 
if the latter is indeed correct.  
The second step is then to consider the 
large $N$ scaling behavior of the 
correlators by making the rescaling $t\rightarrow 
N t$.  

Let us now determine the general form of the 
two-point correlators using the 
relation 
\EQ
\e^{-S_{sugra}[\phi_0]}
=\langle \e^{\int d\tau \sum_I \phi_0^I \cal{O}_I}\rangle
 ,\quad 
\phi^I (x)|_{r=r_b}=\phi^I_0  , \quad  r_b = q^{1/7} 
\EN
where $\tau$ is of course the Wick-rotated 
euclidian time and $I$  labels independent {\it diagonalized} degrees of 
freedom in the spectrum 
of the supergravity fluctuations and 
the corresponding matrix-model operators 
with definite generalized conformal 
dimensions. 
This shows that the connected two-point 
functions are just given by the supergravity 
action evaluated to the second order of the 
boundary value $\phi^I_0$. Since, apart from the overall 
10 dimensional Newton constant $g_s^2\ell_s^8$,  
variable constants ($g_s, N, \ell_s$ ) 
enter in the 
effective action only through 
$q \sim g_sN\ell_s^7$, the generalized 
conformal symmetry is sufficient to fix the 
general form of the two-point functions as 
\EQ
\langle {\cal O}_I (\tau_1) {\cal O}_I (\tau_2)\rangle_c
\sim {1\over g_s^2\ell_s^8}q^{(\Delta_I+6)/5} 
 |\tau_1-\tau_2|^{-(7\Delta_I+12)/5}
\label{2pcorrelator}
\EN 
where we have assumed that the matrix-model 
operators are normalized such that their engineering 
dimensions   with respect to the length is -1, and 
$\Delta_I $ is the generalized conformal 
dimension of the operator; namely 
${\cal O}_I(\tau) \rightarrow 
{\cal O}_I'(\tau') =\lambda^{\Delta_I}{\cal O}_I(\tau), 
\quad \tau'=\lambda^{-1}\tau, \quad 
g_s\rightarrow g_s'=\lambda^3 g_s$ for scaling, and similarly for special conformal transformation.  

 In \cite{sekiyo}, we have performed a complete harmonic 
analysis of the bosonic fluctuations around the D0 background 
and explicitly confirmed the validity of 
the above formula for all 
the bosonic physical fluctuations of supergravity.  
The bosonic fluctuations are described by the modified 
Bessel equation,  
$
\Bigl(
-\partial_z^2 -{1\over z}\partial_z-\partial_{\tau}^2 +{\nu_I^2\over z^2}
\Bigr)\phi^I(z, \tau)=0, 
$
where $z=2q^{1/2}r^{-5/2}/5$ is the `quasi'-Poincar\'{e} 
coordinate and the order of 
the Bessel function is related to the generalized 
conformal dimension by 
$ \Delta_I = -1 +{10\over 7}\nu_I$.  The spectrum of 
the dimensions $\Delta_I$ is classified from the 
11 dimensional viewpoint as 
$\Delta_I = -1 +2n_I + {4\over 7}\ell_I$ where 
$n_I=1 -n_++n_-$ is determined by the kinematical boost 
dimensions $n_{\pm}$, which are nothing but 
the number of upper light-cone indices $\pm$ respectively, 
and $\ell_I$ is the order of the 
harmonics.  The factional dependence on the order of the 
harmonics comes from our normalization of the 
operators such that their engineering dimensions are 
uniformly $-1$, which leads to the harmonic expansion 
in terms of the normalized transverse 
coordinates $\tilde{X}_i=X_i/q^{1/7}$. The results for the  generalized conformal dimensions 
are consistent with the known results \cite{kabattaylor} 
 from the lowest order  perturbative computation 
for the matrix-model operators coupled with 
supergravity fields.  Note that since we are 
just dealing with quantum mechanics 
there is no problem in determining the (generalized) 
conformal dimensions of these operators. 
 A few examples of the operators ${\cal O}(\tau) $ are 
\EQAN
\Delta=-3+{4\ell\over 7}: \, \, T_{\ell, i_1 i_2 \cdots i_{\ell}}^{++}&=&{1\over R}{\rm STr}(\tilde{X}_{i_1}\tilde{X}_{i_2}
\ldots \tilde{X}_{i_{\ell}} +\cdots ) , \quad  (\ell\ge 2)\\
\Delta=-1+{4\ell\over 7}:\, \, T_{\ell,i_1 i_2 \cdots i_{\ell}}^{+i}&=&{1\over R}{\rm STr}(\dot{X}_i
\tilde{X}_{i_1}\tilde{X}_{i_2}
\ldots \tilde{X}_{i_{\ell}} +\cdots ) , \quad (\ell\ge 2)\\
\Delta=+1+{4\ell\over 7}:\, \, T^{ij}_{\ell, i_1 i_2 \cdots i_{\ell}} &=&
{1\over R}{\rm STr}(\dot{X}_i\dot{X}_j
\tilde{X}_{i_1}\tilde{X}_{i_2}
\ldots \tilde{X}_{i_{\ell}} +\cdots ), \quad (\ell\ge 2)
  \quad \\
 etc.&& 
\EQNN
For more details, I would like to invite the reader to  the paper \cite{sekiyo}. 

Let me mention some notable  features of our results 
before discussing the implications for the BFSS conjecture. 
First, the correlators have fractional dependence on 
both $g_s$ and $N$, which can never be reproduced  
from the perturbative computations of the 
correlators. This is not surprising if we recall that such perturbative computations would be plagued by 
infrared divergencies.  Furthermore, the 
angular-momentum independent part of the 
$g_s$-and $N$- dependencies of the 
the dilaton-10D energy-
momentum correlators ($\Delta =1+{4\over 7}\ell$) agree with that of the 
entropy \cite{peetpol} of the nonextremal D0 solution at a given 
temperature $T_H$ :
$ 
S \sim N^2 ( g_sN)^{-3/5}(\ell_sT_H)^{9/5}.
$ 
This provides evidence for the fact that the 
correlator corresponding to the energy-momentum tensor 
without mixing of other modes adequately counts the number of degrees of freedom in the low-energy 
regime of many D-particle dynamics from 10 dimensional 
viewpoint. 

Let us finally discuss the large $N$ scaling behavior of the result (\ref{2pcorrelator})  
and its implications. By making the scaling for the time 
$\tau_1-\tau_2 \rightarrow N(\tau_1-\tau_2)$, we find that the 
correlators scale as $N^{2d_{IMF}}$ with 
\EQ
d_{IMF} = (1+{1\over 5})(n_+-n_- -1) -({1\over 5}+{1\over 7})\ell .
\label{Ndim}
\EN
It is remarkable that except for the additional fraction 
$1/5$ in both the first and the second term in 
(\ref{Ndim}), this is just consistent with the boost 
transformation of the operators.  Note that the 
faction $-1/7$ just accounts for one factor of $N$ 
in $1/q^{1/7}$ coming from the normalization. 
What is not clear is the origin of the anomalous dimensions 
$-1/5$ for the transverse directions and 
$\pm 1/5$ for the upper light-cone indices $\pm$. 
In particular, the large $N$ scaling implies 
a {\it shrinking} behavior $N^{-1/5}$ in the transverse 
direction, which is quite opposite to what we 
naively expect from holography! 

Does this contradict holography? 
Not necessarily. The reason is that 
the large-time correlators of 
the operators with higher partial waves are
 not directly measuring the extension 
of wave functions.  It is conceivable that the 
complicated time-dependent dynamics 
effectively screens the correlation with respect to 
the transverse extension of the wave functions. 
After all, it is very hard to believe that the dynamics 
in the large $N$ IMF could be consistent with 
supergravity, unless the holographic growth of the 
system is somehow screened to become 
an unobservable effect.  
Here we have to recall that the oblique AdS/CFT 
correspondence  on which the above results 
are based has an intrinsic infrared cutoff 
$r < r_B \sim q^{1/7}\propto N^{1/7}$. 
This bound is bigger than the well known 
mean-field estimate $N^{1/9}$. The same estimate  is also 
obtained from a simple counting of 
the degree of freedom for describing the 
Schwarzschild black hole in Matrix theory 
\cite{bfks}. However, 
it is much smaller than the lower bound of 
order $ N^{1/3}$ \cite{polchinski} 
for the eigenvalue distribution which is 
derived using a virial theorem.  
Thus, the oblique AdS/CFT correspondence only enables us to 
 predict the large $N$ behaviors 
of the system put in a box whose size is much smaller 
than the real quantum system, while it is bigger than the 
classical size of the individual objects.  
To discuss the system whose size is 
consistent with the lower bound, we have to renormalize the 
system to bigger sizes. 
This is a difficult dynamical problem. 

Here we may reverse the direction of the arguments. Instead of directly studying the size 
renormalization, we can ask what the size of the system must be if we demand that the large $N$ behavior 
 be consistent with boost invariance.  
Since the consistency with boost invariance requires that 
there should be no anomalous dimensions 
with respect to the  transverse directions, 
let us change the cutoff for the distribution of the 
eigenvalue by making a scaling 
$X_i \rightarrow N^{1/5}X_i$ at  {\it  fixed} $g_s$ and $N$. 
If we assume that the similarity law is 
satisfied for this system, the scaling changes the infrared cutoff from the order $N^{1/7}$ 
to the order $N^{1/5}\times N^{1/7}=N^{(1/3) + \epsilon}, \,  \epsilon = 1/105$.  
This is consistent with and is very close to the lower bound 
$\propto N^{1/3}$.  Although this argument is very naive and it is not 
at all clear whether this procedure  eliminates the anomalous dimensions in the 
light-like directions too,\footnote{The anomalous dimensions in the 
light-like direction might be eliminated by 
simply assuming that the dynamical boost 
should be accompanied with a kinematical 
boost. 
} it might be 
regarded as a signal for  the consistency of the  
large $N$ scaling behavior  with 11 dimensional boost invariance.  We should however  keep in mind also that 
our discussion is limited in the region of weak string coupling 
no matter how $g_sN$ is large.  

Apart from clarifying the above question, there are many future problems which should be 
studied.  To conclude, I enumerate 
a few of them: 
(1) the study of the representation theory of the super 
generalized conformal algebra, (2) the computations 
of three and higher-point functions and their implications, and (3) possible applications to 
the computation of S-matrix in the large $N$ limit. 
 
\ack
I would like to thank W. Taylor and P. Townsend for 
interesting conversations related to this work at the 
Strings'99 Conference. 
The present work  is supported in part 
by Grant-in-Aid for Scientific  Research (No. 09640337) 
and Grant-in-Aid for International Scientific Research 
(Joint Research, No. 10044061) from the Ministry of  Education, Science and Culture.

\end{document}